\documentclass[twocolumn,showpacs,preprintnumbers,aps,prl,amsmath,amssymb]{revtex4}

\usepackage{graphicx}
\usepackage{dcolumn}
\usepackage{bm}

\begin{document}

\title{Electromagnetically induced transparency and dark
fluorescence in a cascade three-level Lithium molecule}
\author{Jianbing Qi$^1$ and A. Marjatta Lyyra$^2$}
\affiliation{$^1$Department of Physics and Astronomy, Penn State
University at Berks, Tulpehocken Road, P.O. Box 7009, Reading, PA
19610}
\affiliation{$^2$Physics Department, Temple University,
Philadelphia, PA 19112}
\date{\today}

\begin{abstract}
We observed electromagnetically induced transparency (EIT) and
dark fluorescence in a cascade three-level diatomic Lithium system
using Optical-Optical Double Resonance (OODR) spectroscopy. When a
strong coupling laser couples the intermediate state
$A^{1}\Sigma^{+}_{u}(v=13, J=14)$ to the upper state
$G^{1}\Pi_{g}(v=11, J=14)$ of $^7Li_2$, the fluorescence from both
$A^{1}\Sigma^{+}_{u}$ and $G^{1}\Pi_{g}$ states was drastically
reduced as the weak probe laser was tuned through the resonance
transition between the ground state $X^{1}\Sigma^{+}_{g}(v=4,
J=15)$ and the excited state $A^{1}\Sigma^{+}_{u}(v=13, J=14)$.
The strong coupling laser makes an optically thick medium
transparent for the probe transition. In addition, The fact that
fluorescence from the upper state $G^{1}\Pi_{g}(v=11, J=14)$ was
also dark when both lasers were tuned at resonance implies that
the molecules were trapped in the ground state. We used density
matrix methods to simulate the response of an open molecular
three-level system to the action of a strong coupling field and a
weak probe field. The analytical solutions were obtained under the
steady-state condition. We have incorporated the magnetic sublevel
(M) degeneracy of the rotational levels in the lineshape analysis
and report $|M|$ dependent lineshape splitting. The theoretical
calculations are in excellent agreement with the observed
fluorescence spectra. We show that the coherence is remarkably
preserved even when the coupling field was detuned far from the
resonance.
\end{abstract}

 \maketitle
 \section{I. Introduction}
Multilevel atomic and molecular systems offer many possibilities
for the investigation of coherence effects and quantum control of
the interactions among the quantum participants. In recent years,
substantial attention has been paid to the study of coherence
effects in atomic and molecular systems ~\cite{1,2,3,4}, such as
coherent population trapping (CPT) ~\cite{5, 6, 7},
electromagnetic induced transparency (EIT) ~\cite{8,9,10,11},
ultraslow propagation of light ~\cite{12,13}, and Autler-Townes
splitting ~\cite{14,15,16,17}. More and more experiments are
shifted from atomic systems to molecular systems for more
practical applications ~\cite{18,19}. The multitude of quantum
levels of molecular systems provides rich coupling schemes and
thus a test ground for the study of coherence effects in molecular
systems. However, compared to atomic systems, molecules have small
transition dipole moments. A general characteristic of molecular
systems is that they have many relaxation pathways. This in turn
makes these systems much more open compared to closed atomic
systems where excited states decay channels are limited.
Furthermore, the degeneracy of the energy levels and other
complications make the observation of coherence effects
considerably more challenging from an experimental point of view.
The Rabi frequency, the key parameter, is proportional to the
transition dipole moment matrix element and the coupling field
amplitude. Thus cw laser experiments that involve small transition
dipole moment matrix elements are therefore quite difficult.
However, a judicious choice of laser wavelengths and beam
propagation geometry can help overcome the Doppler broadening
~\cite{20}. The Autler-Townes splitting was observed in a high
temperature diatomic Lithium gas using multiple resonance
excitation to over come the Doppler effect ~\cite{16,21}.
Recently, EIT in ultracold atomic gases, and Autler-Townes
splitting effect in ultra-cold molecule formation and detection
have been reported ~\cite{22,23,24}. The study of coherence
effects in molecular systems is timely and important not only for
fundamental understanding of these effects, but also for the
practical applications. In this paper, we present the detailed
experimental investigation and the corresponding theoretical
analysis of electromagnetically induced transparency and dark
fluorescence in a cascade three-level diatomic Lithium in an
inhomogeneously broadened environment. We have incorporated the
effect of the magnetic sublevel (M) degeneracy of the rotational
levels in the lineshape analysis and report $|M|$-dependent
lineshape splitting. We show that the coherence is remarkably
preserved even when the coupling field was detuned far from the
resonance. The open property of molecular systems will be
discussed in our theoretical calculation. We also demonstrate that
the coupling laser field dependent splitting of the upper level
can be used as a new method for measuring the molecular transition
dipole moment matrix element ~\cite{21}.

The paper is organized as follows. In section II, we present the
theoretical model and the derivation of the analytical expressions
to account for the experiments. We describe the experimental
observations in section III. The discussion of the theoretical
calculations using the experimental parameters is given in section
IV. Finally, a summary is presented.
\maketitle
\section{II. Theoretical Framework}
\subsection{A. Density Matrix Equation of Motion.}
The excitation scheme for a three-level molecular system
interacting with two laser fields is indicated in Fig. 1(a). We
consider a moving molecule situated in a travelling wave
$\vec{E}_i{(}z,t{)}=\vec{e}_{i}E_{i}\cos{(k_{i}z-\omega_{i}t)}$.
The Hamiltonian, $\emph{H}$, is given by,
\begin{equation}
    H = H_{0}+H_{int},\label{1}
\end{equation}
    where
\begin{equation}
H_{0} = \sum_{i=1}^{3} {\varepsilon_{i}|i><i|}\label{2}
\end{equation} is the
molecular Hamiltonian, and $\varepsilon_{i}$ is the energy
eigenvalue of the isolated molecule in state $|i>$. We assume
$\varepsilon_{1}=0$ for simplicity and all other states are
measured relative to state $|1>$. The
\begin{equation}
H_{int}=\sum_{i\neq{j}} <i|(-\vec{\mu}\cdot\vec{E}|j>) =
-\sum_{i\neq{j}}\mu_{ij}E_{i}\label{3}
\end{equation}
is the dipole interaction Hamiltonian, and $\mu_{ij}$ is the
transition dipole moment for a molecule undergoing
$|i,v',J'>\leftrightarrow|j,v,J>$ transition. The evolution of the
molecular density matrix for a molecule moving with velocity
\emph{v} is governed by the master equation ~\cite{25},
\begin{equation}
\frac{\partial\varrho}{\partial{t}}+\vec{v}.\nabla\varrho=-\frac{i}{\hbar}[H,\varrho]
+{(}\frac{\partial\varrho}{\partial{t}}{)}_{inc},\label{4}
\end{equation}
 where the second term on the left hand-side  represents the damping due
 to spontaneous emission and other irreversible processes.
\begin{figure}
\centering \vskip -4 mm
\includegraphics[width=8 cm]{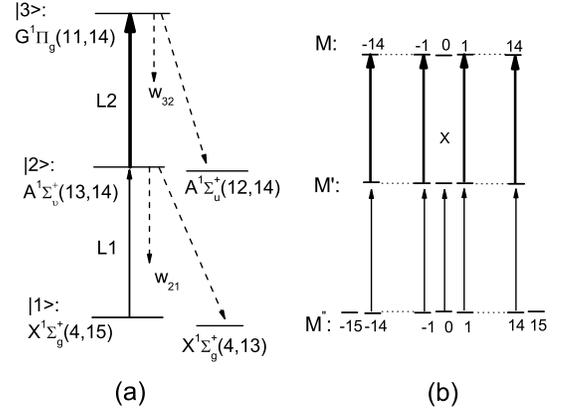}
\caption{$^7Li_2$ three-level cascade scheme: (a) The weak probe
laser, L1 (15642.636 $cm^{-1})$, was used to excite molecules from
the ground state level $X^1\Sigma^{+}_{g}(v=4, J=15)$ to an
excited intermediate level $A^1\Sigma^{+}_{u}(v=13,J=14)$. The
laser, L2 (17053.954 $cm^{-1})$, coupled the intermediate level to
a higher electronic state level $G^{1}\Pi_{g}(v=11, J=14)$. The
fluorescence from $A^{1}\Sigma^{+}_{u}(v=13,J=14)$ to
$X^1\Sigma^{+}_{g}(v=4, J=13)$ and $G^{1}\Pi_{g}$ to
$A^{1}\Sigma^{+}_{u}(v=12,J=14)$ were monitored. (b) The coupling
details of the magnetic sublevels (M=-J,-J-1,...,J-1,J) of (a):
For linearly polarized light, the selection rules require
$\Delta{M}=0$, thus The $M" =\pm{15}$ of the ground state levels
are decoupled from the first transition(P transition: $\Delta
J=-1$), while there is no $M=0 \rightarrow M'=0$ coupling for the
upper transition (Q transition: $\Delta J=0$).}
\end{figure}

Before we apply the density matrix formalism to interpret the
experimental results, we have to consider the relaxation details
of the level system in Fig. 1(a). The first laser L1 excites the
$^7Li_{2}$ molecules from the ground state level
$X^1\Sigma^{+}_{g}(v=4, J=15)$ (level $|1>$) to the intermediate
level $A^{1}\Sigma^{+}_{u}(v=13, J=14)$ (level $|2>$), and the
second laser L2 couples the $A^{1}\Sigma^{+}_{u}(v=13, J=14)$ to
the upper excited state $G^{1}\Pi_{g}(v=11, J=14)$ (level $|3>$).
Molecules in any specific rovibrational level of an excited
electronic state can decay to many other rovibrational levels of
lower electronic states, and only part of them decay back to their
initial state. The upper excited electronic state $G^1\Pi_{g}$ can
decay to two lower electronic states of $B^{1}\Pi_{u}$ and
$A^{1}\Sigma^+_u$. The $A^{1}\Sigma^{+}_{u}$ state is the first
singlet excited electronic state of the Lithium molecule.
Molecules in a rovibrational level of the $A^{1}\Sigma^{+}_{u}$
state can decay to vibrational levels of the ground electronic
state $X^{1}\Sigma^{+}_{g}$. In the sense of the description of
the total decay rate of level $|3>$ to other energy levels, there
is no difference between $B^{1}\Pi_{u}$ and $A^{1}\Sigma^{+}_{u}$
state.

We assume that the total radiative decay rate of the excited
states $|2>$ and $|3>$ are $\gamma_{2}$ and $\gamma_{3}$,
respectively. The branching ratios $b_{2}$ and $b_{3}$ stand for
the percentage of molecules in the level $|2>$ and the level $|3>$
that decay back to the ground-state $|1>$ and level $|2>$,
respectively. When the branching ratios are equal to unity the
three-level system is closed. The laser frequency detunings for a
stationary molecule are defined as
\begin{equation}
\delta_{1}=\omega_{1}-\omega_{21},
\end{equation}
 and
\begin{equation}
\delta_{2}=\omega_{2}-\omega_{32},
\end{equation}
where
$\omega_{ij}=\frac{\varepsilon_{i}-\varepsilon_{j}}{\hbar}$ is the
resonance transition frequency between $|i>$ and $|j>$. The Rabi
frequency of the corresponding laser field is defined as:
\begin{equation}
g_{i}=\mu_{ij}E_{i}/\hbar.
\end{equation}
We assume that the population of the ground state $|1>$ has been
replenished at the rate $\Lambda$, and only the ground-state is
replenished. The laser beam has a finite beam size and therefore
the transverse motion of molecules can remove molecules from the
interaction region before decay. This will introduce an effective
additional relaxation of the excited states and the ground state.
In order to account for this transit time, we simulate it with an
effective decay rate $w(w<<\gamma_{i})$ for all populations and
polarizations. Then, the explicit form of equation (4) is
\begin{eqnarray}
(\frac{\partial}{\partial{t}}+v_{z}\frac{\partial}{\partial{z}})\varrho_{33}=
        ig_{2}\cos(k_{2}z-\omega_{2}t)(\varrho_{32}-\varrho_{23})\\
            -(\gamma_{3}+\emph{w})\varrho_{33},\nonumber\label{equ.8}
\end{eqnarray}
\begin{eqnarray}
(\frac{\partial}{\partial{t}}+v_{z}\frac{\partial}{\partial{z}})\varrho_{22}=
            -ig_{2}\cos(k_{2}z-\omega_{2}t)(\varrho_{32}-\varrho_{23})\\
            -ig_{1}\cos(k_{1}z-\omega_{1}t)(\varrho_{12}-\varrho_{21})\nonumber\\
            +W_{32}\varrho_{33}-(\gamma_{2}+\emph{w})\varrho_{22},\nonumber\label{equ.9}
\end{eqnarray}
\begin{eqnarray}
(\frac{\partial}{\partial{t}}+v_{z}\frac{\partial}{\partial{z}})\varrho_{11}=
            ig_{1}\cos(k_{1}z-\omega_{1}t)(\varrho_{12}-\varrho_{21})\\
            \Lambda+W_{21}\varrho_{22}-\emph{w}\varrho_{11},\nonumber\label{equ.10}
\end{eqnarray}
\begin{eqnarray}
(\frac{\partial}{\partial{t}}+v_{z}\frac{\partial}{\partial{z}})\varrho_{32}=
            ig_{2}\cos(k_{2}z-\omega_{2}t)(\varrho_{33}-\varrho_{22})\\
                +(-i\omega_{32}-\gamma_{32}-\emph{w})\varrho_{32}\nonumber\\
                +ig_{1}\cos(k_{1}z-\omega_{1}t)\varrho_{31}\nonumber,\nonumber\label{equ.11}
\end{eqnarray}
\begin{eqnarray}
(\frac{\partial}{\partial{t}}+v_{z}\frac{\partial}{\partial{z}})\varrho_{31}=
        [-i\omega_{31}-(\gamma_{31}+\emph{w})]\varrho_{31}\\
        -ig_{2}\cos(k_{2}z-\omega_{2}t)\varrho_{21}\nonumber\\
        +ig_{1}\cos(k_{1}z-\omega_{1}t)\varrho_{32},\nonumber\label{equ.12}
\end{eqnarray}
\begin{eqnarray}
(\frac{\partial}{\partial{t}}+v_{z}\frac{\partial}{\partial{z}})\varrho_{21}=
        [-i\omega_{21}-(\gamma_{21}+\emph{w})]\varrho_{21}\\
        +ig_{1}\cos(k_{1}z-\omega_{1}t)(\varrho_{22}-\varrho_{11})\nonumber\\
        -ig_{2}\cos(k_{2}z-\omega_{2}t)\varrho_{31},\nonumber\label{equ.13}
\end{eqnarray}
where the $W_{ij}$ is the population decay rate from level $|i>$
to $|j>$, $W_{32}=b_{3}\gamma_{3}$, and $W_{21}=b_{2}\gamma_{2}$,
and $\gamma_{ij}^{c}$ represents the collisional dephasing rate.
The polarization decay rate $\gamma_{ij}$ is given by
\begin{equation}
\gamma_{ij}=\gamma_{ji}=\frac{1}{2}\sum_{k}(W_{ik}+W_{ki}).\nonumber
\end{equation}
Let
\begin{equation}
\Delta_{1}=\omega_{1}-\omega_{12}-k_{1}v_{z}=\delta_{1}-k_{1}v_{z},\label{equ.14}
\end{equation}
and
\begin{equation}
    \Delta_{2}=\omega_{2}-\omega_{23}-k_{2}v_{z}=\delta_{2}-k_{2}v_{z},\label{equ.15}
\end{equation}
where $v_{z}$ is the velocity component of the molecule in the
laser propagation direction. Equations (8)-(13) can be changed
into ones for the density-matrix elements of the slowly varying
function of time and space by setting
\begin{eqnarray}
\varrho_{21}=\rho_{21}e^{i(k_{1}z-\omega_{1}t)},\label{equ.16}
\end{eqnarray}
\begin{eqnarray}
\varrho_{32}=\rho_{32}e^{i(k_{2}z-\omega_{2}t)},\label{equ.17}
\end{eqnarray}
\begin{eqnarray}
\varrho_{31}=\rho_{31}e^{i[(k_{1}+k_{2})z-(\omega_{1}+\omega_{2})t]}.\label{equ.18}
\end{eqnarray}
After applying the rotating wave approximation, the above
equations (8)-(13) can be written as:
\begin{eqnarray}
    \frac{d\rho_{33}}{d{t}}=
        i\frac{g_{2}}{2}(\rho_{32}-\rho_{23})
        -(\gamma_{3}+\emph{w})\rho_{33},\label{equ.19}
\end{eqnarray}
\begin{eqnarray}
    \frac{d\rho_{22}}{d{t}}=
        i\frac{g_{1}}{2}(\rho_{21}-\rho_{12})
        -i\frac{g_{2}}{2}(\rho_{32}-\rho_{23})\\
        -(\gamma_{2}+\emph{w})\rho_{22}
        +W_{32}\rho_{33},\nonumber\label{equ.20}
\end{eqnarray}
\begin{eqnarray}
\frac{d\rho_{11}}{d{t}}=
    i\frac{g_{1}}{2}(\rho_{12}-\rho_{21})
    +W_{21}\rho_{22}
        -\emph{w}\rho_{11}
            +\Lambda,\label{equ.21}
\end{eqnarray}
\begin{eqnarray}
\frac{d\rho_{32}}{d{t}}=
    i\frac{g_{2}}{2}(\rho_{33}-\rho_{22})
    +i\frac{g_{1}}{2}\rho_{31}
    +i\Delta_{2}\rho_{32}\\
    -(\gamma_{23}+\emph{w})\rho_{32},\nonumber\label{equ.22}
\end{eqnarray}
\begin{eqnarray}
\frac{d\rho_{31}}{dt}=
    i\frac{g_{1}}{2}\rho_{32}
    -i\frac{g_{2}}{2}\rho_{21}
    -(\gamma_{13}+\emph{w})\rho_{31}\\
    +i(\Delta_{1}+\Delta_{2})\rho_{31},\nonumber\label{equ.23}
\end{eqnarray}
\begin{eqnarray}
\frac{d\rho_{21}}{d{t}}=
    i\frac{g_{1}}{2}(\rho_{22}-\rho_{11})
    -i\frac{g_{2}}{2}\rho_{31}
    +i\Delta_{1}\rho_{21}\\
    -(\gamma_{12}+\emph{w})\rho_{21},\nonumber\label{equ.24}
\end{eqnarray}

In the steady-state limit, we can solve the above equations
iteratively for the population ($\rho_{ii}$) of each level to the
lowest order of the weak probe laser Rabi frequency $g_{1}$, but
to all orders in $g_{2}$. After some lengthy algebra, we obtain
the non-normalized analytical solutions for the populations of the
two excited states:
\begin{widetext}
\begin{eqnarray}
    \rho_{22}=
        -\frac{{g_{1}^{2}\rho_{11}^{(0)}}}{2D(\Delta_{2})}
        Im\left\{
        \frac{
        \frac{g_{2}^{2}}{4}
        (1-\frac{W_{32}}{\gamma_{3}+\emph{w}})
        [\Delta_{2}-i(\gamma_{32}+\emph{w})]
        +A[\Delta_{1}+\Delta_{2}+i(\gamma_{31}+\emph{w})]}
        {[\Delta_{1}+i(\gamma_{21}+\emph{w})]
        [\Delta_{1}+\Delta_{2}+i(\gamma_{31}+\emph{w})]
        -\frac{g_{2}^{2}}{4}}
        \right\},\label{equ.25}
\end{eqnarray}
and
\begin{eqnarray}
    \rho_{33}=
        \frac{{g_{1}^{2}{g_{2}^{2}}\rho_{11}^{(0)}}}
        {8D(\Delta_{2})(\gamma_{3}+\emph{w})}
        Im\left\{
        \frac{
        -2(\gamma_{32}+\emph{w})
        [\Delta_{1}+\Delta_{2}+i(\gamma_{31}+\emph{w})]
        +(\gamma_{2}+\emph{w})
        [\Delta_{2}-i(\gamma_{32}+\emph{w})]}
                {[\Delta_{1}+i(\gamma_{21}+\emph{w})]
        [\Delta_{1}+\Delta_{2}+i(\gamma_{31}+\emph{w})]
        -\frac{g_{2}^{2}}{4}}
        \right\},\label{equ.26}
\end{eqnarray}
\end{widetext}
where
\begin{equation}
    A=\Delta_{2}^{2}+(\gamma_{32}+\emph{w})^{2}
        +\frac{g_{2}^{2}(\gamma_{32}+\emph{w})}
         {2(\gamma_{3}+\emph{w})},\nonumber
\end{equation}
\begin{equation}
    D(\Delta_{2})=
    A(\gamma_{2}+\emph{w})
    +\frac{g_{2}^{2}(\gamma_{23}+\emph{w})}{2}
    (1-\frac{W_{32}}{\gamma_{3}+\emph{w}}),\nonumber
\end{equation}
and
    $\rho_{11}^{(0)}=\frac{\Lambda}{\emph{w}}$
is the initial population without the probe laser field. We can
see that the system will be ideally closed if
$W_{32}=\gamma_{3}+\emph{w}$, and the expressions will be greatly
simplified.
\subsection{B. Doppler Effect}
Let us assume that two laser beams counter propagate along the z
axis, the probe laser travels to the right (positive), and the
coupling laser to the left (negative). Due to the Doppler effect,
a molecule moving with a positive velocity $\emph{v}_{z}$ with
respect to the rest frame will $\emph{see}$ the probe laser and
coupling laser frequencies $\omega_{1}$, and $\omega_{2}$,
respectively, as:
\begin{equation}
\omega_{1}(v_{z})=\omega_{1}-\frac{v_{z}}{c}\omega_{1},\label{27}
\end{equation}
and
\begin{equation}
\omega_{2}(v_{z})=\omega_{2}+\frac{v_{z}}{c}\omega_{2}.\label{equ.28}
\end{equation}
 We define the velocity dependent detunings as
\begin{equation}
\Delta_{1}(\emph{v}_{z})=\delta_{1}-\frac{\emph{v}_{z}}{c}\omega_{1},\label{equ.29}
\end{equation}
and
\begin{equation}
\Delta_{2}(\emph{v}_{z})=\delta_{2}+\frac{\emph{v}_{z}}{c}\omega_{2}.\label{equ.30}
\end{equation}
The velocity dependent laser detunings can be expressed in laser
frequency detuning and the transition frequency as follow:
\begin{equation}
\Delta_{1}(\emph{v}_{z})=(1-\frac{\emph{v}_{z}}{c})\delta_{1}
            -\frac{\emph{v}_{z}}{c}\omega_{12},\label{equ.31}
\end{equation}and
\begin{equation}
\Delta_{2}(\emph{v}_{z})=(1+\frac{\emph{v}_{z}}{c})\delta_{2}
        +\frac{\emph{v}_{z}}{c}\omega_{23}.\label{equ.32}
\end{equation}
At thermal equilibrium, the molecules in a gas phase follow the
Maxwellian velocity distribution, in one dimension, which is given
as ~\cite{26}:
\begin{eqnarray}
N(\emph{v}_{z})=\frac{1}{\sqrt{\pi}u_{p}}\exp(-\frac{\emph{v}_{z}^{2}}{u_{p}^2}),\label{equ.33}
\end{eqnarray}
where $u_{p}=(\frac{2kT}{m})^{1/2}$ is the most probable velocity
of the molecules, k is the Boltzmann's constant, m is the mass of
a molecule, and T is the temperature. The experimental
observations should be the sum $\rho_{ii}$ for all velocity
groups.
\begin{equation}
\langle\rho_{ii}\rangle_{Doppler}=
    \int_{-\infty}^{+\infty}\rho_{ii}N(\emph{v}_{z})d\emph{v}_{z}.\label{equ.34}
\end{equation}
\begin{figure}
\centering \vskip -2 mm
\includegraphics[width=8 cm]{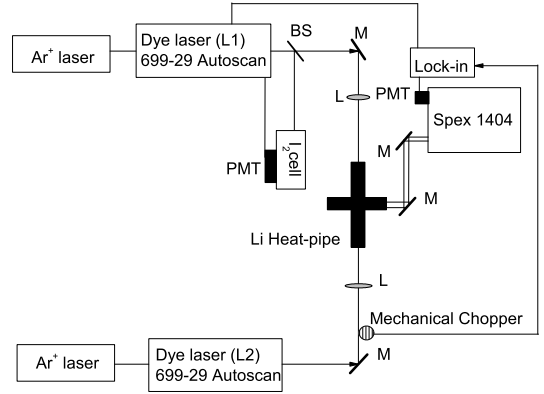}
\caption{\protect\label{Fig.2}Experimental set-up:
Two linearly polarized counter propagating laser beams were
aligned coaxially, and were focused at the center of the Lithium
heat-pipe. The fluorescence was collected and focused to the
monochromator (SPEX 1404) from the side window. The signal was
amplified by the lock-in amplifier and the output was recorded on
the Coherent 699-29 Autoscan computer. (M: Mirror, BS: Beam
splitter, L: Lens, PMT: Photo-multiplier)}
\end{figure}
\subsection{C. $|M|$-Dependent Rabi Frequency}
For each rotational angular momentum \emph{J}, there are
\emph{2J+1} magnetic sublevels, \emph{M=-J,-(J-1),...,J-1,J},
which specify the projection of the total angular momentum J along
a laboratory fixed Y-axis. The interaction of each magnetic
sublevel with the laser field depends not only on the transition
(P, Q, or R) but also on the polarization of the laser field
~\cite{27, 28}. The Rabi frequency $g_{i}$ for each laser field
and for a given molecular transition of $(v', J')\leftarrow(v, J)$
can be written in the form:
\begin{eqnarray}
g_{i}=\mu_{ij}E_{i}/\hbar=<v'|\mu_{e}|v>f(J'JM'M;\Lambda'\Lambda)E_{i}/\hbar,\label{equ.35}
\end{eqnarray}
where $\mu_{e}$ is the electronic transition dipole moment,
$\mu_{e}=<\Lambda'|\mu|\Lambda>$, $f(J'JM'M;\Lambda'\Lambda )$ is
the rotational line strength factor for transition
$|J'M'\Lambda'>\leftrightarrow|JM\Lambda>$, and $E_{i}$ is the
laser field strength. For a linearly polarized laser field, the
rotational line strength factor for the $Q(\Delta J=J-J'=0)$
transition is
\begin{eqnarray}
    f_{Q}=\frac{|M|}{\sqrt{J(J+1)}},\label{equ.36}
\end{eqnarray}
and for the P$(\Delta J=J-J'=-1)$ transition,
\begin{eqnarray}
    f_{P}=\sqrt{\frac{(J^{2}-M^{2})}{(2J+1)(2J-1)}}.\label{equ.37}
\end{eqnarray}
For linearly polarized light, the transition selection rules
require $\Delta{M}=M'-M=0$, thus there is no $M=0 \rightarrow
M'=0$ coupling for upper transition. The $M''=\pm{15}$ of the
ground state are decoupled from the first transition also. The
coupling of the three-level configuration of the Fig. 1(a) can be
viewed as 28 M-dependent couplings for L2 and 29 couplings for L1
, as shown in Fig. 1(b). The Rabi frequency depends on the
absolute value of the magnetic sublevel $|M|$. This results in
$|M|$-dependent population expressions of equation (25) and (26)
also. The observed fluorescence signal, apart from a
proportionality factor, can be calculated by integrating the
$\rho_{ii}$ over the velocity distribution and summing over all
$|M|$, i.e.
\begin{eqnarray}
\rho_{ii}(\delta_{1}, \delta_{2})\propto
    \sum_{|M|}\int_{-\infty}^{+\infty}\rho_{ii}N(\emph{v}_{z})d\emph{v}_{z}.\label{38}
\end{eqnarray}
\begin{figure}
\centering \vskip -8 mm
\includegraphics[width=8 cm]{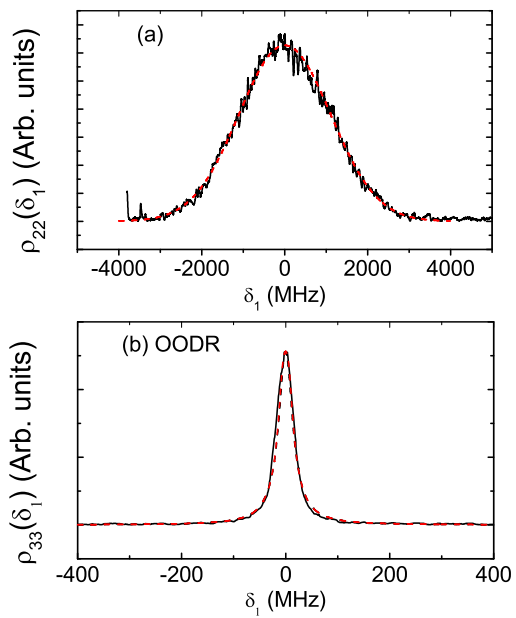}
\caption{\protect\label{Fig.3} (color online) (a)
The Doppler broadened fluorescence spectrum of $A^1\Sigma^+_u(13,
14)$ to $X^1\Sigma^+_g(4,13)$ is plotted as a function of the
detuning of the probe laser without the coupling laser. (b)
Measured OODR signal of $G^1\Pi_g(11, 14)$ along with the
calculation with 1 mW coupling laser power. (Solid lines:
experiment, dashed lines: calculation.)}
\end{figure}
\begin{figure}
\centering \vskip -8 mm
\includegraphics[width=8 cm]{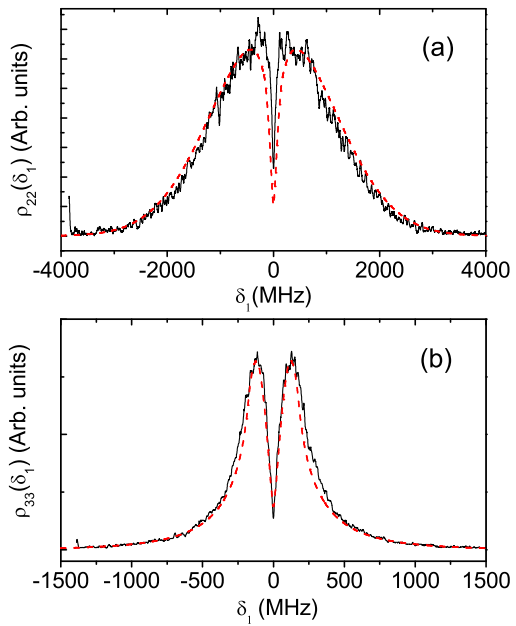}
\caption{\protect\label{Fig.4} (color online)
Measured experimental spectra along with the calculations. (a)
Fluorescence from level $|2>$. (b) Fluorescence from level $|3>$.
The fitting parameters are: $<v'|\mu_{e}|v>= 1.45(\pm0.1) a.u.$,
$\gamma_{13}^{c}=\gamma_{23}^{c}=1 MHz$, $\gamma_{12}^{c}$=5 MHz.
The coupling laser power is 480 mW. (Solid lines: experimental,
dashed lines: calculation.)}
\end{figure}
\begin{figure}
\centering \vskip -8 mm
\includegraphics[clip, width=8 cm]{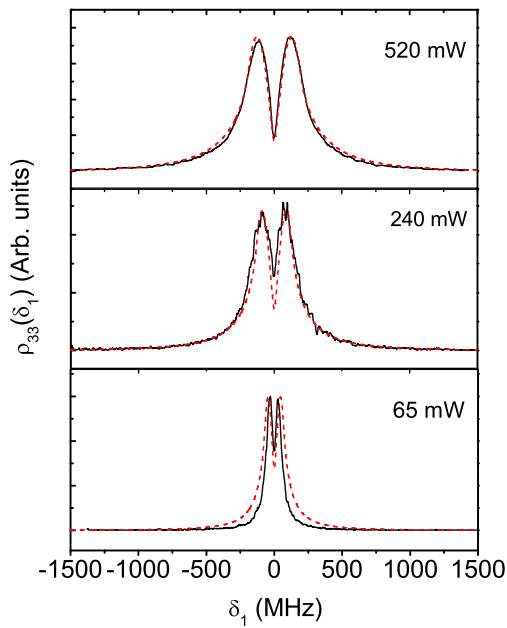}
\caption{\protect\label{Fig.5} (color online) Measured
fluorescence spectra of the upper level as a function of the probe
laser detuning ($\delta_{1}$) for different coupling laser power
with the coupling laser frequency tuned at resonance
($\delta_{2}=0$). The fitting parameters are the same as in Fig.
4: $<v'|\mu_{e}|v>= 1.45(\pm{0.1}) a.u.$,
$\gamma_{13}^{c}=\gamma_{23}^{c}=1 MHz$, $\gamma_{12}^{c}$=5 MHz.
(solid lines: experimental, dashed lines: calculation)}
\end{figure}
\begin{figure}
\centering \vskip -11 mm
\includegraphics[width=8 cm]{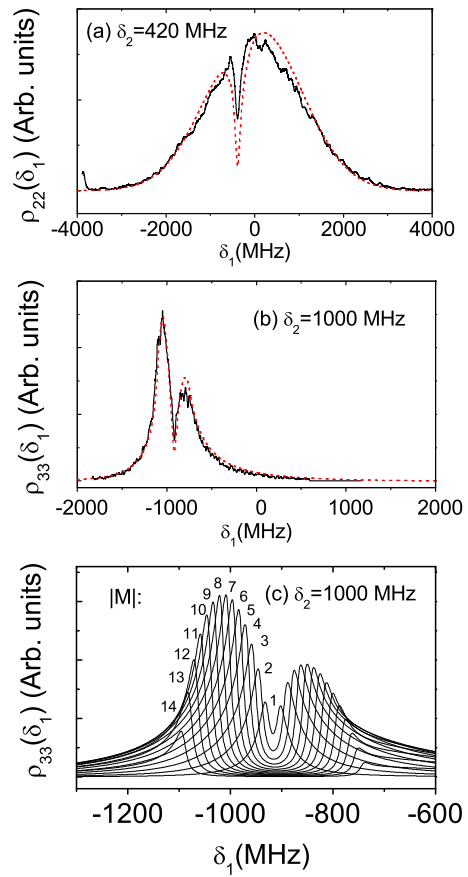}
\caption{\protect\label{Fig.6}(color online) Fluorescence spectra
for the coupling laser detuned from the resonance frequency: The
coupling laser power is same as that in Fig. 4. (a) The coupling
laser is detuned at $\delta_2=420 MHz$ above the resonance
frequency. The EIT dip of level $|2>$ shifted by 385 MHz below the
resonance frequency. (b) The coupling laser is detuned 1.0 GHz
above the resonance frequency. The splitting of the upper state
$|3>$ is still preserved, but shifted 917 MHz below the resonance
frequency. (Solid lines: experiment, dashed lines: calculation)
(c) Calculated 14 $|M|$-components of (b) before summation using
Eq. (38).}
\end{figure}
\section{III. Experimental Results}
The experimental set-up is shown in Fig. 2. This is a typical
Optical-Optical Double Resonance (OODR) scheme. Lithium molecules
are generated in a five-arm stainless steel oven with the
temperature around 1000 Kelvin, and with Argon buffer gas pressure
around 100-300 mTorr. Two Coherent 699-29 Autoscan dye lasers were
used to produce the required laser wavelengths. Two linearly
polarized laser beams were arranged in counter-propagating
configuration and aligned coaxially.

We monitored the population of the intermediate state
$A^1\Sigma^+_u(v=13, J=14)$ (level $|2>$) by detecting its
fluorescence to the ground-state rovibrational level
$X^1\Sigma^+_g(v=4, J=13)$. The corresponding wavelength is
6377.83 \AA\ in air. The population of the upper state
$G^{1}\Pi_{g}(v=11, J=14)$ (level $|3>$) was monitored by
detecting its fluorescence to the $A^{1}\Sigma^{+}_{u}(v=12,J=14)$
state with the wavelength of 5791.30 \AA\ in air. The fluorescence
was collected and focused to the monochromator (SPEX 1404) through
a set of mirrors from the side window of the heat-pipe oven. The
selected fluorescence was detected by a cooled
photomultiplier(PMT) at the exit slit of the SPEX when the
monochromator was set to the corresponding spontaneous emission
wavelength. The PMT signal was amplified by a lock-in amplifier(SR
850), and the output was recorded on the 699-29 Autoscan computer
while the probe laser (L1) frequency was scanned. All laser
frequencies were calibrated to $\pm{0.002} cm^{-1}$ with the
standard Iodine spectra ~\cite{29}\@. The first transition from
the ground state level of $^{7}Li_{2}X^{1}\Sigma^{+}_{g}(v=4,
J=15)$ to the excited state $A^{1}\Sigma^{+}_{u}(v=13, J=14)$ is
driven by the probe laser L1, while the transition from
$A^{1}\Sigma^{+}_{u}(v=13, J=14)$ to the upper state level
$G^{1}\Pi_{g}(v=11, J=14)$ is driven by laser L2. In the absence
of laser L2, a frequency scan of the probe laser L1 yields the
usual Doppler broadened fluorescence spectrum of the
$A^{1}\Sigma^{+}_{u}(v=13, J=14)$ as shown in the Fig. 3(a). If
the coupling laser L2 is weak and set at the resonance transition
of the $A^{1}\Sigma^{+}_{u}(v=13,J=14)$ to $G^{1}\Pi_{g}(v=11,
J=14)$, by monitoring the fluorescence of $G^{1}\Pi_{g}(v=11,
J=14)$ to $A^{1}\Sigma^{+}_{u}(v=12, J=14)$, a scan of the probe
laser from $X^{1}\Sigma^{+}_{g}(v=4, J=15)$ to
$A^{1}\Sigma^{+}_{u}(v=13, J=14)$ leads to the usual OODR spectrum
for the upper state, $G^{1}\Pi_{g}(v=11, J=14)$, as shown in Fig.
3(b). Upon increasing the coupling laser power, a sharp dip
emerges at the center of the Doppler broadened fluorescence
spectrum of the $A^{1}\Sigma^{+}_{u}(v=13, J=14)$, as shown in
Fig. 4(a). One may naively interpret the emergence of that dip as
the consequence of the additional transfer of population to the
upper level $G^{1}\Pi_{g}(v=11, J=14)$ by the strong coupling
laser. However, a sharp dip also appears in the middle of the OODR
fluorescence signal of the upper $G^{1}\Pi_{g}(v=11, J=14)$ level.
The OODR fluorescence peak, splits into two components as shown in
Fig. 4(b). The fluorescence of both excited states is drastically
reduced under the action of the strong coupling laser (L2).
Because the intensity of the fluorescence is proportional to the
population of the corresponding excited state, the origin of the
fluorescence dips is based on the fact that the molecules can not
be excited by the probe laser and the coupling laser to either the
intermediate level $A^{1}\Sigma^{+}_{u}(v=13, J=14)$ or the upper
level $G^{1}\Pi_{g}(v=11, J=14)$ from the ground-state under a
strong coupling laser. Since the strong coupling laser L2 modified
the transition from $X^{1}\Sigma^{+}_{g}(v=4, J=15)$ to
$A^{1}\Sigma^{+}_{u}(v=13, J=14)$, the ground state molecules can
not absorb the probe laser photons and be excited to the excited
state $A^{1}\Sigma^{+}_{u}(v=13, J=14)$ at the resonance
frequency. The remarkable result is that the second transition
does not transfer the molecules to the higher excited
$G^{1}\Pi_{g}$ state either. The experimental results also show
that the stronger the coupling laser is, the deeper and wider are
the dips as shown in Fig. 5. In a sense of the first transition,
the molecule becomes transparent under the action of the strong
coupling laser (electromagnetically induced transparency) and thus
the molecules must stay in the ground state.
\section{IV. Discussion}
In order to carry out a comparison between the experimental
spectra and the theory, we calculate Eq. (38) based on Eq. (25)
and Eq. (26) using the experimental data for the transition
frequencies $\omega_{21}$ and $\omega_{32}$. The lifetimes of
level $|2>$ $(\tau_{2}= 1/\gamma_{2})$ and $|3>$
$(\tau_{3}=1/\gamma_{3})$ were based on the references
~\cite{30,31}. These values are 18 ns, and 16.15 ns, respectively.
From Fig. 3(a) (the probe laser scan) we obtained the most
probable molecular velocity by measuring the Doppler line-width,
which is 2.6 GHz. The coupling beam waist ($1/e^{2}$) is 360
$\mu$m. The week probe laser beam is 222 $\mu$m ($\sim$1 mW). The
transit rate (\emph{w}) of the molecules entering and leaving the
interaction region can be estimated according to reference
~\cite{32} and is $\sim$ 2 MHz. The branching ratios $b_{2}$ and
$b_{3}$ can be estimated from the Franck Condon factor calculation
to be 0.1 and 0.2, respectively. We perform the calculations based
on the analytical solution of Eq. (26) by searching the value of
the transition dipole moment matrix element and the
$\gamma_{ij}^{c}$ to best match the experimental spectrum of
$\rho_{33}$ in Fig. 4(b). The resulting value of the transition
dipole moment matrix element $<v'|\mu_{e}|v>$ for
$G^{1}\Pi_{g}-A^{1}\Sigma^{+}_{u}$ is 1.45 ($\pm{0.1}$) a.u..
After the completion of this step, we calculate the corresponding
spectrum of $\rho_{22}$, which is again in good agreement with the
experimental spectrum shown in Fig. 4(a) as dashed lines. However,
the experimental dip is much narrower than the theoretical
calculation. Both the theory and the experimental spectra clearly
show that a strong coupling laser modified the transitions. The
molecules stay in the ground state even though the laser (L1) was
tuned to the resonance of the first transition as long as the
coupling laser (L2) couples the upper transition with adequate
coupling strength $(g_{2})$. The optically opaque molecular gas
now becomes transparent for laser L1 (EIT). Again, the dip is not
due to the population transfer to the upper state $|3>$ by the
coupling laser L2, because the upper state has no population
either. The fluorescence of both excited states becomes dark in
the presence of the strong coupling laser. Keeping all parameters
fixed, decreasing the strength of the coupling laser, we obtain
the single peak OODR for $\rho_{33}$ as shown by the dashed lines
in Fig. 3(b). The calculated $\rho_{22}$ without the coupling
laser is identical to the Doppler broadened profile as shown in
Fig. 3(a)\@.

As indicated in the equation (25) and (26) that the Rabi frequency
of the coupling field $g_{2}$, has a dominant influence on the
depth and width of the dips of $\rho_{33}$ once
$g_{2}^{2}\gg4(\gamma_{21}+\emph{w})(\gamma_{31}+\emph{w})$. The
decay rate of the upper level $|3>$ and the branching ratios
$b_{i}$, have a contribution to the depth of the dip of the
spectra as well. The branching ratio $b_{i}$, the collision rate
$\gamma_{ij}^{c}$ and the transit rate $\emph{w}$ have a dominant
contribution to the linewidth and the wings of the upper state
spectra. This is understandable and expected compared to a closed
system. A large $\gamma_{ij}^{c}$ means that the coherence will be
destroyed quickly, and a large transit rate \emph{w} implies an
effective shorter lifetime of the excited levels, while small
values of \emph{w} and large branching ratios imply that the
system is better described by a closed system. The dip of
$\rho_{22}$ is none zero since the P transition of the probe laser
can populate the M=0 sublevel of $A^1\Sigma_{u}^{+}$ state, while
the coupling field transition is a Q transition, the M=0 level is
decoupled from the coupling field transition(see Fig. 1(b)). Also,
the Doppler broadening greatly reduces the width of the
transparency window.

When the coupling laser is off resonance, the dips are still
preserved as long as the coupling field is strong enough. However,
the position of the dips will change to the opposite side of the
detuning of the coupling field ($\delta_{2}$). We can easily find,
by checking the integral equation (34), that the position of the
dip is at the modified two-photon transition: $\delta_{1}
=-|\frac{k_{1}}{k_{2}}|\delta_{2}$ due to the Doppler effect, and
at $\delta_{1}=-\delta_{2}$ for Doppler free case. This is a
completely coherent process, since the detuning of L2 prevents the
population buildup on level $|3>$. The splitting of this component
depends on the coupling field strength and the detuning of
$\delta_{2}$ as well as the linewidths of the two excited states.
Two experimental spectra with coupling laser detuned from the
resonance by 420 MHz, and 1.0 GHz, respectively, but with the same
laser intensity as in the Fig. 4 are shown in Fig. 6(a)-(b), which
show that the coherence is robustly preserved. We plot 14 $|M|$
sublevel components of Fig. 6(b) in an expanded scale to show the
$|M|$-dependent splitting by using equations (26), and (34) in
Fig. 6(c). For a Q transition the splitting of each $|M|$
component is proportional to the value of $|M|$\@.
\section{V. Summary}
In summary, we have observed the electromagnetically induced
transparency (EIT) and dark fluorescence in an inhomogeneous
broadened Lithium molecular system. The power dependent upper
state splitting spectrum provides a useful method to
experimentally measure the transition dipole moment matrix
element. The value of this parameter from fits of the experimental
spectra agrees very well with the theoretical calculation. It
could provide new insights into the electronic structures and
dynamics of Rydberg states ~\cite{21}. In the fitting of the
experimental spectra we find that the branching ratio value can be
varied over a large range and still give a reasonable fit. It
implies that it is possible to observe EIT in a very open system,
such as pre-dissociated molecular states. We demonstrated that the
coherence was remarkably preserved even when the coupling field
was detuned far from the resonance. We have discussed a systematic
approach to the treatment of the response of a three-level open
molecular system to the presence of two laser fields. The
theoretical model and the treatment of the degeneracy of the
rotational levels agree very well with the experimental spectra.
\section{Acknowledgements}
We thank Profs. L. M. Narducci and F. C. Spano for their valuable
discussions, as well as A. Lazoudis, T. Kirova and J. Magnes for
their technical help in the lab. J. Qi is grateful for the summer
research support from Penn State University at Berks. We
acknowledge support from NSF grants PHY0245311 and PHY9983533 to
Temple University .

\end{document}